# 3D photo-responsive optical devices manufactured by advanced printing technologies

Adam Szukalski,[a] Sureeporn Uttiya,[a] Francesca D'Elia,[b] Alberto Portone,[a] Dario Pisignano,[a,c] Luana Persano,[a] Andrea Camposeo*[a]

[a]NEST, Istituto Nanoscienze-CNR, Piazza S. Silvestro 12, I-56127 Pisa, Italy; [b]NEST, Scuola Normale Superiore, Piazza San Silvestro 12, I-56127 Pisa, Italy; [c]Dipartimento di Fisica, Università di Pisa, Largo B. Pontecorvo 3, I-56127 Pisa, Italy

**ABSTRACT**

Photonic components responsive to external optical stimuli are attracting increasing interest, because their properties can be manipulated by light with fast switching times, high spatial definition, and potentially remote control. These aspects can be further enhanced by novel architectures, which have been recently enabled by the availability of 3D printing and additive manufacturing technologies. However, current methods are still limited to passive optical materials, whereas photo-responsive materials would require the development of 3D printing techniques able to preserve the optical properties of photoactive compounds and to achieve high spatial resolution to precisely control the propagation of light. Also, optical losses in 3D printed materials are an issue to be addressed. Here we report on advanced additive manufacturing technologies, specifically designed to embed photo-responsive compounds in 3D optical devices. The properties of 3D printed devices can be controlled by external UV and visible light beams, with characteristic switching times in the range 1-10 s.

## 1. INTRODUCTION

A paradigm shift is currently occurring in the design and fabrication of optical and photonic devices, especially as a consequence of the introduction of additive manufacturing (AM) technologies.[1] Conventional photonic and optoelectronic components (lenses, optical fibers, light sources and detectors) typically feature very simple geometries, such as curved spherical surfaces, filaments, and planar multilayers, possibly ordered in arrays to build the final devices. Recent examples include, for instance, arrays of Mach-Zender interferometers, which are designed and assembled in cascade to perform machine learning tasks.[2] The current progress of 3-dimensional (3D) printing and AM technologies is pushing optical and photonic devices towards new, unexplored architectures,[3,4] enabled by the realization of 3D material configurations with almost no restriction in terms of printed geometry. AM technologies comprise a variety of fabrication processes which have been classified in seven macro-areas,[5] taking in account their basic operating principles. The starting point in all AM technologies is a digital model of the component to be realized, that is sliced in various layers. The object is then realized by generating consecutively the various layers, in the so-called layer-by-layer fashion, either by curing liquid resins, or by extrusion of melted polymers and viscous solutions, or by sintering fine powders, or by delaminating solid layers.[6] One of the most popular 3D printing methods is the fused deposition modeling (FDM), that is based on the extrusion of a melted polymer filament through a nozzle. The 3D object is built layer-by-layer by translating the nozzle along the 3D path that is given by the sliced digital model.[7] The FDM can be applied to thermoplastic polymers and composites, it can be used to build large-area components and functional devices,[8,9] but it has a spatial resolution limited by the nozzle size (typically of the order of a few hundreds of micrometers). This is one of the reasons why FDM has been mainly used for optical components in long wavelength ranges (terahertz and microwaves).[10] Also relevant in this framework is the often weak optical transmittance of parts printed by FDM at visible wavelengths. In fact, for some applications, such as lab-on-chips[11] and device packaging,[12] FDM would be highly advantageous, because it would allow plastic components with moderate optical transmittance (>50%) to be printed in a fast and low-cost way, and then embedded in functional devices such as supports or windows for optical access to diagnostic environments. This requires, however, the engineering of FDM to comply with transparent thermoplastic materials, such as polycarbonate (PC), polystyrene, and poly(methyl methacrylate). This field is almost unexplored, since most of the research works on FDM-printing with such polymers has been focused on the investigation of the morphological and mechanical characterization of printed parts.[13,14]

*andrea.camposeo@nano.cnr.it; phone +39 050 509517.

Another AM technology for printing optical components is the laser stereolithography (STL),[15] a method based on the photo-polymerization of a liquid resin by means of a laser beam, which is scanned to solidify the individual layers forming the 3D object. The various layers are printed consecutively by moving vertically the object holder. STL has better spatial resolution than FDM, and it may reach the sub-µm scale when performed on materials with nonlinear absorption phenomena,[16] though at the cost of slower printing rate. Nevertheless, even standard UV STL has been used for successfully printing various optical components, such as aspheric lenses, mirrors and optical guides operating in the visible range.[17-19] In this framework, a current challenge is to develop photo-curable materials and printing methods, that are suitable to achieve optically-active components, namely systems which can feature some specific optical properties (i.e. emission at well-defined spectral bands, nonlinear and time-changing properties, etc.). Indeed, photonic devices and integrated systems are continuously shifting from static and passive systems, basically able to control the propagation of light,[2,20] to active elements capable of reconfiguring themselves in response to external signals.[21] This can be currently obtained by phase-changing materials, whose crystalline structure and physical properties can be reversibly modified by irradiation with laser pulses.[22,23] Other examples include the exploitation of stimuli-responsive molecular compounds,[24] which can be incorporated in photonic devices and employed for tuning the emission wavelength of organic lasers,[25,26] for making birefringent films and nanofibers,[27,28] and for controlling the optical transmittance of smart optical windows.[29,30] To fully exploit the enormous potential of such novel optical materials, the 3D printing technology has to be addressed to manufacturing with them, without degrading their properties.

Here we report on printing processes for the fabrication of transparent layers by FDM and STL, and photo-responsive layers by photo-polymerization. By optimization of the process parameters (single-layer thickness and printing speed), PC optical windows are realized by FDM, with optical transmittance at visible wavelengths larger than 50%. In addition, optical windows with high transparency (>80%) are printed by STL. We also show the possibility to functionalize photo-curable layers by a photo-responsive molecular compound. In particular, we succeeded in printing photoactive layers, suitable for reversible control of the intensity of light propagating through the printed components. This allows the intensity of a polarized light beam to be controlled by a second light beam, with characteristic switching times of the order of 1-10 s. These results open interesting perspectives for utilizing 3D printed, transparent and photoactive components in a wide range of devices, including analytical and sensing optical tools.

## 2. METHODS

A PC filament (Roboze) was used for printing optical windows by FDM. The E-Shell® 600 (ES600, EnvisionTEC) was used as matrix for samples printed by STL. The printable photo-responsive resin was obtained by mixing the *N*-Ethyl-*N*-(2-hydroxyethyl)-4-(4-nitrophenylazo)aniline (DR1, Sigma Aldrich) with the ES600 matrix (1% DR1:ES600 weight:weight ratio).

FDM experiments were performed by the ONE+400 system (Roboze), equipped with two independent extruders, with maximum operational temperature of 450 °C. The maximum printing volume is $20 \times 20 \times 20$ cm$^3$ ($X \times Y \times Z$, here $Z$ is the direction parallel to the sample thickness). Samples were printed on a PC substrate, that was kept at a constant temperature of 80 °C. The fabrication of the optical windows was performed by using an extruder with a nozzle of 400 µm diameter. The extruder temperature was varied in a range of 240-290 °C. A temperature of 250 °C was selected as the one allowing samples with higher optical transparency to be printed.

The Andromeda (Sharebot) 3D printer was used for STL experiments. This is a laser 3D printer, with a maximum printing volume of $25 \times 25 \times 25$ cm$^3$. Optical windows were printed by a laser power fluence of 8 mW/cm$^2$ and a layer thickness of 100 µm. The first two layers were printed at a laser scanning rate of $1.5 \times 10^5$ µm/s, whereas the remaining layers were printed using a higher rate ($2.5 \times 10^5$ µm/s). This set of parameters allows a good adhesion to the sample holder to be achieved, as well as a final printed structure with geometrical features in agreement with the original design. After the printing process, the printed part was washed in isopropanol for 5 minutes, and dried under a nitrogen flow. Post-processing surface finishing was performed by coating the samples with a thin and homogenous film of liquid resin (thickness 200 µm), that is then polymerized by exposure to UV light (maximum power fluence 10 mW/cm$^2$) for two minutes. The finished part was washed again for 5 minutes in isopropyl alcohol to remove the uncured layer in contact with air, because oxygen inhibits radical polymerization and leaves the outermost layer uncured.

UV-visible (UV-Vis) transmittance spectra were measured by using either a Lambda 950 (Perkin Elmer) or a V-550 (Jasco) spectrophotometer. Photo-induced birefringence was investigated by a pump-probe experimental set-up, which is typically used for the investigation of the optical Kerr effect.[28] A laser probe beam ($\lambda_{probe}$ = 638 nm, linearly polarized) is sent through the sample at normal incidence. The probe beam polarization is determined by a polarizer positioned between the probe laser source and the sample, while the light transmitted by the sample is analyzed by a second polarizer with axis perpendicular to the first. The intensity of the probe beam transmitted by the sample and by the analyzer is measured by a Si photodiode. In order to induce optical anisotropy in the investigated sample, a pump laser beam ($\lambda_{pump}$ = 532 nm, linearly polarized) irradiates the same area probed by the red laser. The directions of the polarization of the pump and probe laser beams form an angle of 45 degrees. This configuration is known to maximize the photo-induced optical anisotropy.[28] With the pump beam switched off, the sample does not feature any optical birefringence and, as a consequence of the presence of two polarizers with crossed axis along the path of the probe beam, no signal is detected by the photodiode. By switching the pump beam on, an optical anisotropy is induced in the area irradiated by the pump, which leads to a rotation of the polarization of the probe beam passing through the sample, and an increase of the signal transmitted at the second polarizer according to the relation:

$$\frac{I}{I_0} = \sin^2\left(\frac{\pi |\Delta n| d}{\lambda_{probe}}\right) \quad (1)$$

where $d$ is the thickness of the photoactive layer, $\Delta n$ is photo-induced birefringence, and $I_0$ and $I$ are the incident and transmitted probe intensities, respectively. By measuring the temporal evolution of this signal, one is able to measure the characteristic times of the molecular photo-alignment occurring in the printed samples. In order to characterize the possibility to modulate the probe beam by switching on and off the pump, a mechanical chopper with variable modulation frequency in the range 1-800 Hz was positioned along the path of the pump beam.

## 3. RESULTS AND DISCUSSION

Figure 1 summarizes the measured UV-Vis optical properties of squared optical windows made of PC and printed by FDM. An example of the printed sample is shown in the inset of Figure 1b, which evidences the good optical transparency, allowing for easily seeing through the printed component. All the investigated samples show a low optical transmittance in the range 200-300 nm (T=$I_T/I_{in}$<4%, where $I_T$ and $I_{in}$ are the intensities of the transmitted and incident beam, respectively). This is typical of PC, which is, indeed, used as plastic blocking material for UV light.[31] We have investigated optical properties of the printed components upon varying two process parameters, i.e. the thickness of single printed layer (Figure 1a,b) and the printing speed (Figure 1c,d). The intensity of the light transmitted by the samples is found to drop by almost an order of magnitude upon increasing the layer thickness (Figure 1b). This trend is found for wavelengths across the whole visible and near-infrared (NIR) spectral range (400-800 nm), with highest values of the transmittance (T≈50 %) obtained for a layer thickness of 50 µm. In addition, the intensity of transmitted light increases by an order of magnitude upon increasing the printing speed (Figure 1d). Other experimental reports evidenced that in samples printed by FDM, more rough surfaces are associated to higher thicknesses of the single layers and to lower values of the printing speed.[32] The increase of the roughness of samples is expected to increase the intensity of light that is diffused at angles out of the initial propagation direction, and, consequently to decrease the overall optical transmittance. Controlling sintering of adjacent printed filaments[7] and reducing the roughness of the printed structures turns out to be especially critical for improving optical transmittance of layers printed by FDM.

Optical windows printed by STL share some of these features (Figure 2a,b). Here, the intensity of the light transmitted by as-printed samples is found to be in the range 40-50 % in the visible and NIR spectral range. However, post-printing surface finishing is highly effective in improving the transmittance of these samples, leading to T values larger than 80%. The deposition of a thin resin film on printed samples decreases the surface roughness, and the associated diffusion of incident light.[18] Overall, the results here shown evidence that samples with acceptable optical transmittance can be printed by both FDM and STL. In order to achieve values of T comparable to plastic and glass optical windows, a proper surface finishing method has to be applied.

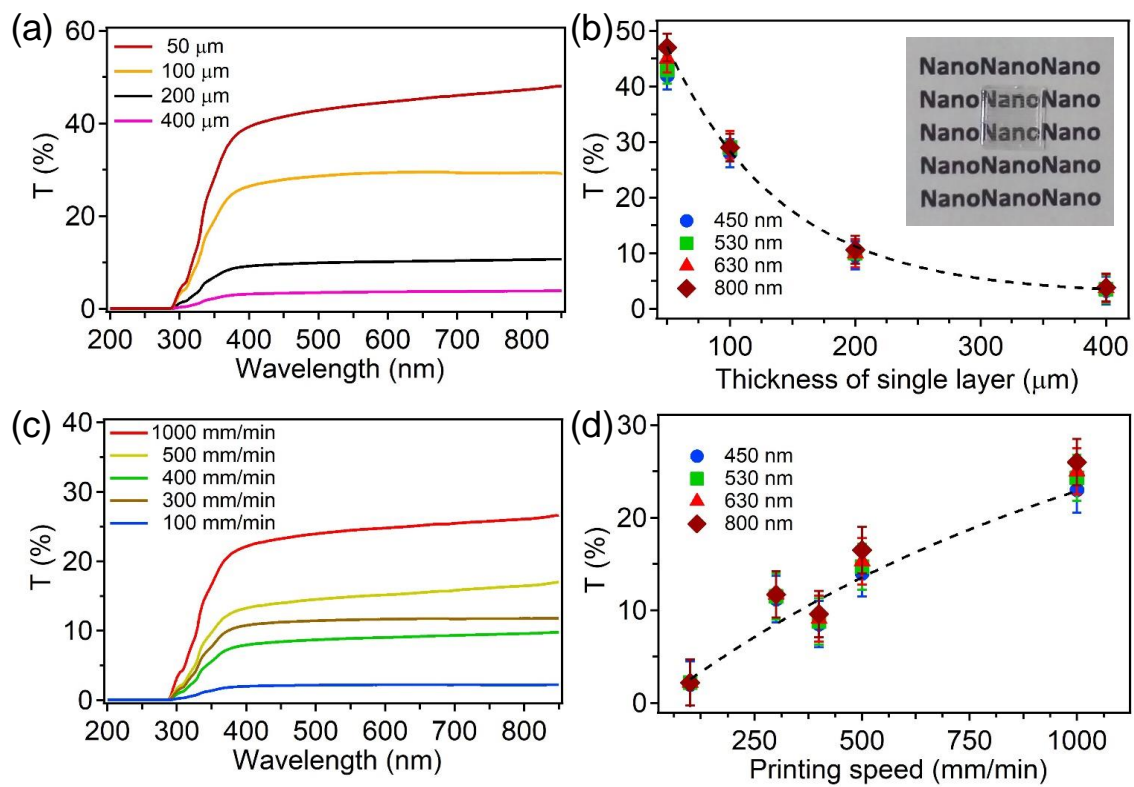

Figure 1. (a) Spectra of the light intensity transmitted (T) by samples made of PS and printed by FDM, for various thicknesses of the single layers. The values of T at fixed wavelength *vs.* the thickness of single printed layers is shown in (b). The dashed line in (b) is a guide for the eye. Inset: photograph of a PC sample printed by FDM, highlighting achieved transparency. Scale bar: 1 cm. (c) Spectral dependence of T on the printing speed. The corresponding values of T at fixed wavelength as a function of the printing speed is shown in (d). The dashed line in (d) is a guide for the eye. The samples used for measurements shown in (a)-(d) have area of 1×1 cm$^2$ and thickness of 1 mm.

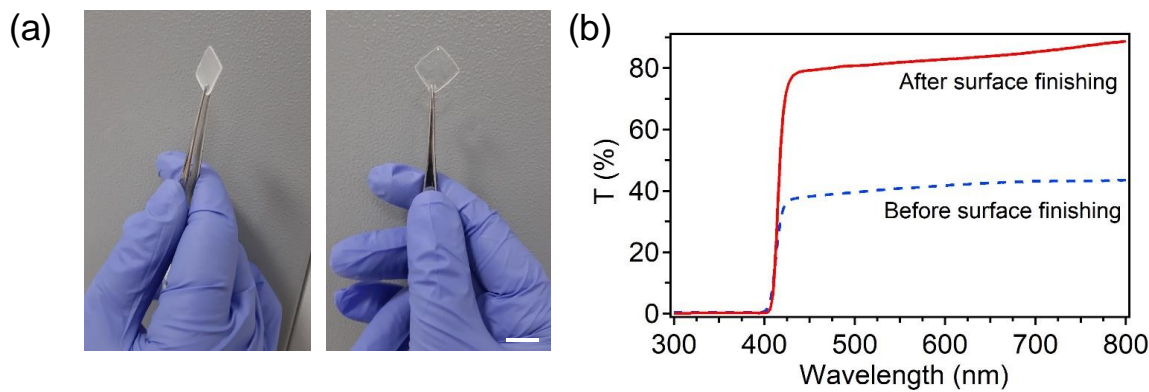

Figure 2. (a) Photographs of the optical windows printed by STL before (left image) and after (right image) the surface finishing process. Scale bar: 1 cm. (b) UV-Vis spectra of T for samples printed by STL before (blue dashed line) and after (red continuous line) the surface treatment. The samples used for the measurements are those shown in (a).

To explore the possibility of expanding the range of optical materials to photoactive systems, which can be shaped by 3D printing, we investigated the properties of UV-polymerized layers containing photorefractive compounds. These layers are made by spin-coating the E600/DR1 mixture on glass substrates, and curing them under UV light (10 mW/cm$^2$) in a nitrogen environment. In such systems, the embedded photo-isomerizable molecules undergo a series of *trans-cis-trans* cycles, which leads to the formation of an anisotropic distribution of the molecule orientation, driven by the angular redistribution of molecules due to the *trans-cis* conversion and rotational diffusion coming from thermal agitation.[27,33] These effects are responsible of the build-up of optical anisotropy (dichroism and birefringence) in layers containing azobenzene derivatives upon optical excitation with wavelength resonant with their absorption band. Figure 3a shows the time evolution of the intensity of the probe beam transmitted by the sample and by crossed polarizers. The initial zero signal increases in presence of the pump beam, reaching a saturation on timescales of hundreds of seconds. The all-optical control of the probe is completely reversible, as shown in Figure 3a, highlighting a decay of the probe beam intensity by turning off the pump. The signal build-up and decay is well described by a double exponential function, with time constants of about 2 s for the fast component and 30-50 s for the slow one.

The optical anisotropy occurring in photo-responsive, nonlinear chromophores in polymer matrices has been investigated in various host-guest systems, and most of these studies emphasized a complex dynamics characterizing both the build-up and the decay regimes.[27,34-38] A bi-exponential function is often used to account for the temporal evolution of the observed rise and decay of optical nonlinearities.[27,36] More specifically, for the relaxation process occurring when light is turned off, the fast component is attributed to thermodynamically-activated *cis*-to-*trans* conformational changes, while the slow one originates from angular re-distribution of the *trans* isomers, and the mobility and relaxation of the polymer chains of the host. In other works,[37,38] a stretched exponential function was found to describe the dynamics of the photo-induced nonlinear properties, and a tight relationship between the mechanical/rheological properties of the polymeric hosts and the temporal evolution of the photo-induced alignment of the guest molecules was also evidenced. As 3D printed structures are going to be developed by embedding photoactive dopants in thermoplastic and photo-curable polymers, the interplay between the matrix and the guest molecule will play a fundamental role for either enhancing or depleting the ultimate nonlinear optical properties. This requires further experimental and theoretical investigation of such effects, for rationalizing occurring phenomena and for engineering 3D printing processes, which can produce complex structures with nonlinear optical properties, hopefully outperforming those observed in planar films.

Finally, the possibility of modulating light by light, namely the modulation of the probe beam by turning on and off the pump by a mechanical chopper is shown in Figure 3b. Data show that the probe beam can be modulated at frequencies of the order of hundreds of Hz, as observed typically for azobenzene derivatives in polymer and biopolymer matrices.[35] The temporal profile of the modulated intensity of the probe signal reflects the complex dynamics of the photo-induced alignment of the active molecules. Indeed, while the pump beam is switched on and off with ms characteristic times, the probe signal follows an exponential trend with much longer characteristic times, as expected from the results shown in Figure 3a. Recently, pyrazoline derivatives have been introduced which showing much faster photoswitching could be used for enhancing the modulation frequencies of printed structures.[28] Overall, the photoactive and photo-curable resins here studied show optical properties suitable for all-optical control of light beams, that is promising for future exploitation in 3D printing processes by STL.

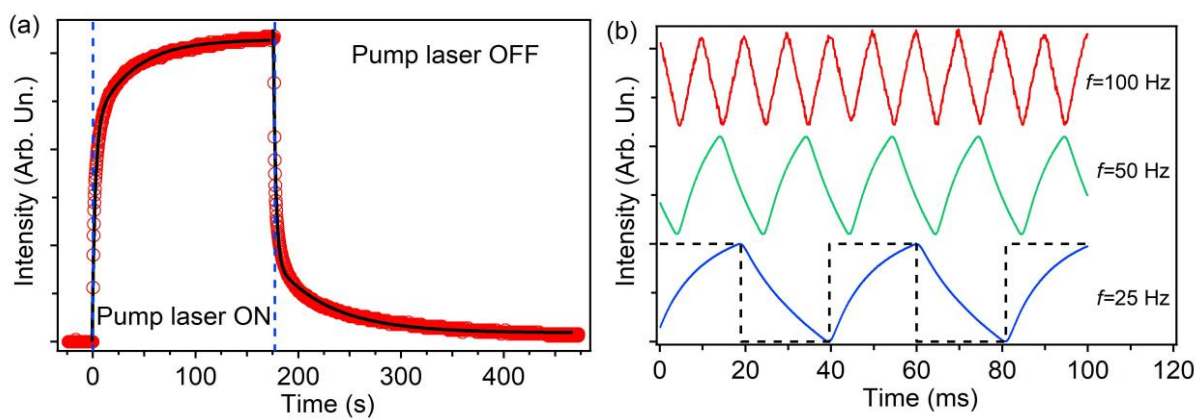

Figure 3. (a) Temporal evolution of the intensity of the light transmitted by a layer of ES600 doped with DR1, upon switching the pump laser beam with emission at 532 nm. The interval with pump laser 'on' is marked by vertical dashed lines. The continuous lines are fit to the data by bi-exponential functions. (b) Examples of time evolution of a light beam passing through a layer of ES600/DR1 and modulated at 25, 50 and 100 Hz by switching the pump beam (dashed line).

## 4. CONCLUSIONS

In summary, optical windows have been printed by FDM and STL, and studied in their transparency. The investigation of the dependence of the optical properties on printing parameters evidence a tight relationship. In particular, high printing speeds and low layer thicknesses are suitable to increase the optical transmittance of structures printed by FDM, while surface finishing by deposition of a film of uncured resin allows highly transparent sheets to be obtained by STL. In addition, photo-curable sheets with light-responsive optical properties have been demonstrated, which allows the intensity of a light beam to be modulated by an external beam at frequencies of hundreds of Hz. Such results are relevant for the development of a novel class of 3D printed optical devices, whose properties can be varied in real-time by control light beams.

*Acknowledgments*. The research leading to these results has received funding from the European Research Council (ERC) under the European Union's Horizon 2020 research and innovation programme (grant agreement No. 682157, "xPRINT").